\documentclass{aa}  
\usepackage{graphicx}
\usepackage{tabularx}
\usepackage{color}
\usepackage{txfonts}
\newcommand{\gp}[1]{{\color{black}#1}}
\newcommand{\gpp}[1]{{\color{black}#1}}
\begin{document} 
\title{ALMA study of the HD~100453~AB system and the tidal interaction of the companion with the disk}

\titlerunning{HD~100453 with ALMA}

   \author{G.  van der Plas \inst{1}
        \and F.  M\'enard \inst{1}
        \and J.-F.  Gonzalez \inst{2}
        \and S.  Perez \inst{3,4}
        \and L.  Rodet \inst{1}
        \and C.  Pinte \inst{1,5}
        \and L.  Cieza \inst{6}
        \and S.  Casassus \inst{3,4}
        \and M.  Benisty \inst{1,4}
        }

   \institute{Univ. Grenoble Alpes, CNRS, IPAG (UMR 5274), F-38000 Grenoble, France
        \and Univ Lyon, Univ Lyon1, Ens de Lyon, CNRS, Centre de Recherche Astrophysique de Lyon UMR5574, F-69230, Saint-Genis-Laval, France
        \and Millenium Nucleus Protoplanetary Disks in ALMA Early Science, Universidad de Chile, Casilla 36-D, Santiago, Chile
        \and Departamento de Astronomia, Universidad de Chile, Casilla 36-D, Santiago, Chile
        \and Monash Centre for Astrophysics (MoCA) and School of Physics and Astronomy, Monash University, Clayton Vic 3800, Australia
        \and Nucleo de Astronomia, Facultad de Ingenieria, Universidad Diego Portales, Av Ejercito 441, Santiago, Chile
        }

   \date{Received ...; accepted ...}

  \abstract
   {The complex system HD~100453~AB with a ring-like circumprimary disk and two spiral arms, one of which is pointing to the secondary, is a good laboratory to test spiral formation theories.}
   {To resolve the dust and gas distribution in the disk around HD~100453~A and to quantify the interaction of HD~100453~B with the circumprimary disk.}
   {Using ALMA band 6 dust continuum and CO isotopologue observations we study the HD~100453~AB system with a spatial resolution of 0\farcs09 $\times$ 0\farcs17 at 234~GHz. We use SPH simulations and orbital fitting to investigate the tidal influence of the companion on the disk.}
   {We resolve the continuum emission around HD~100453~A into a disk between 0\farcs22 and 0\farcs40 with an inclination of 29.5\degr ~and a position angle of 151.0\degr, an unresolved inner disk, and \gp{excess mm emission cospatial with the northern spiral arm which was previously detected using scattered light observations.}  We also detect CO emission from 7 au (well within the disk cavity) out to 1\farcs10, i.e., overlapping with HD~100453~B at least in projection. The outer CO disk PA and inclination differ by up to 10\degr~from the values found for the inner CO disk and the dust continuum emission, which we interpret as due to gravitational interaction with \gp{HD~100453~B}. Both the spatial extent of the CO disk and the detection of mm emission \gp{at the same location as }the northern spiral arm are in disagreement with the previously proposed near co-planar orbit of HD~100453~B.}
   {We conclude that HD~100453~B has an orbit that is significantly misaligned with the circumprimary disk. Because it is unclear whether such an orbit can explain the observed system geometry we highlight an alternative scenario that explains all detected disk features where \gp{another, } (yet) undetected, low mass close companion within the disk cavity, shepherds a misaligned inner disk whose slowly precessing shadows excite the spiral arms.}


   \maketitle
%
\section{Introduction}
    
    Protoplanetary (PP) disks are a natural byproduct of star formation. These disks dissipate with a typical timescale of 2 to 3 million years \citep[see e.g. the review by][and references therein]{2011ARA&A..49...67W} and planet formation during the evolution and dissipation of the disk appears to be the rule rather than the exception \citep[e.g,][]{2015ApJ...807...45D}. The mechanisms that allow the gas and small dust grains in the disk to coalesce into planetary systems are not clear yet and high angular resolution studies of PP disks are necessary to solve this part of the planet formation puzzle.

    Our current best tools to study PP disks at high spatial resolution are (sub-)mm interferometers such as ALMA and extreme AO high-contrast imagers such as the Gemini Planet Imager \citep[Gemini/GPI][]{2014PNAS..11112661M} and the Spectro-Polarimetric High-contrast Exoplanet REsearch \citep[VLT/SPHERE][]{2008SPIE.7014E..18B}. Each of them now routinely yields spatial resolutions below 0\farcs1 but each traces different regions of the disks. The scattered light traces the small $\approx$ micron sized dust grains high up in the disk surface, while the longer wavelength observations can trace both the larger, typically mm sized, dust grains in the disk mid plane, as well as the intermediate disk layers through many different molecular gas lines. 

    As we observe PP disks at increasingly high spatial resolution it becomes clear that substructures in these disks are common, and that understanding these substructures is essential to understand disk evolution and planet formation. The most common structures found so far are [1] opacity cavities ranging between a few to over 100 au that sometimes contain a small misaligned inner disk \citep[such as, e.g., HD~142527, see ][]{2015ApJ...798L..44M}, \gp{ where this disk also casts a shadow on the outer disk \citep{2012ApJ...754L..31C}, } [2] (multiple) rings and / or cavities \gp{\citep[e.g. ][]{2016ApJ...820L..40A,2018ApJ...863...44A}}, [3] large spiral arms \citep[such as i.e. HD~142527, see ][]{2014ApJ...785L..12C} \gp{, or HD~100453, see \cite{2015ApJ...813L...2W}}, and [4] azimuthal dust concentrations with various contrast often interpreted as dust trapping in vortices \citep[such as i.e. IRS~48 and HD~34282, see ][]{2013Sci...340.1199V,2017A&A...607A..55V}. All of these features can be induced by the gravitational interaction with a forming body (e.g., a planet) but also by other processes that don't require a massive body within the disk such as snow lines \citep{2006ApJ...640.1115L,2017A&A...600A.140S}, a pressure gradient at the edge of a dead zone \citep{1999ApJ...513..805L}, self-induced dust traps \citep{2013Natur.499..184L, 2017MNRAS.467.1984G}, stellar fly-by \citep{2005AJ....129.2481Q}, and others. Studying these features using different proxies narrows down their possible origins and thus helps building a list of processes that are dominant in disk dispersal and planet formation. The nearby HD~100453~AB system is an ideal candidate for such a study.

    {\bf HD~100453~A} at 103$^{+3}_{-4}$ pc \citep{2016A&A...595A...1G} is an A9Ve star with an age of 10$\pm$2 Myr and a mass of 1.7 M$_\odot$ \citep{2009ApJ...697..557C}. It is orbited by a companion (hereafter called HD~100453~B) that was first noticed by \citet{2006A&A...445..331C} and later confirmed to be co-moving by \citet{2009ApJ...697..557C}.  The spectral type of HD~100453~B was estimated to be between M4V and M4.5V with a mass of 0.2 $\pm$ 0.04 M$_{\odot}$ \citep{2009ApJ...697..557C}. \gp{\citet{2018ApJ...854..130W} recently published new astrometric measurements further confirming the bound nature of the companion orbit, and placing it at a projected separation of $\approx$ 1\farcs05 from the primary at a position angle of 131.95\degr at the time of the observations we present in this manuscript.}
    
     The disk surrounding HD~100453~A is highly structured  and complex. There is little material accreting onto the central star with an upper limit to the accretion rate of 1.4 $\times$ 10$^{-9}$ M$_\odot$ yr$^{-1}$ \citep{2009ApJ...697..557C}. \citet{2015ApJ...813L...2W} resolved a disk cavity and an outer disk between 0\farcs18 and 0\farcs25 in radius, as well as two nearly symmetric spiral arms extending out to r=38 au (distances scaled to a distance of 103 au), and \citet{2017A&A...597A..42B} saw two symmetric shadows on the outer disk, all in scattered light. Similar features have been detected in other transition disks \gp{\citep[objects whose inner disk regions have undergone substantial clearing, see e.g. ][]{2014prpl.conf..497E}} such as the ones around HD~135344~B  \citep{2016A&A...595A.113S} and  HD~142527 \citep{2015ApJ...798L..44M}. In these cases the shadow cast by a small, misaligned, inner disk was deemed responsible.  NIR infrared interferometric observations have indeed detected such a misaligned inner disk around HD~100453~A with a half light radius of $\approx$ 1 au \citep{2015A&A...581A.107M, 2017A&A...599A..85L}, and \citet{2017A&A...604L..10M} calculate a position angle and inclination for the inner disk ($i$ = 45\degr, PA = 82\degr) and for the outer disk ($i$ = -38\degr, PA = 142\degr) using the assumption that the shadows are cast by the inner disk. Finally, \citet{2003A&A...402..767M} report an unresolved detection of the disk at 1.2 mm with 265 $\pm$ 21 mJy, and \citet{2018ApJ...854..130W} use part of the data we present here to determine a counter-clockwise rotation direction for the disk.  The grand design spiral arm structure in this system has been connected to the companion by \citet{2016ApJ...816L..12D}, who used hydrodynamical and radiative transfer simulations to show that a close-to-coplanar orbit of the companion can explain the main disk features detected in scattered light assuming the disk is oriented close to face-on.\\

    In this paper we present high angular resolution ALMA band 6 observations of the HD~100453 system to measure the dust and gas distribution in the disk (Sections \ref{sec:obs} and \ref{sec:res}). We use hydrodynamical SPH and radiative transfer models to investigate whether our observations are consistent with the previously suggested coplanar companion as origin for the spiral arms (Section \ref{sec:ana}), and we discuss our results in Section \ref{sec:discussion}. We conclude in Section \ref{sec:conclusion} that this is unlikely to be the case and offer an alternative scenario to explain the system geometry where an as of yet undetected companion inside the disk cavity drives a slowly precessing misaligned inner disk whose shadow cast on the outer disk triggers the spiral arms. 

\section{Observations and data reduction}\label{sec:obs}

    ALMA Early Science Cycle 3 observations were conducted in a compact configuration on April 23$^{\mathrm{rd}}$ 2016 with 13.1 minutes of total time on-source and in an extended configuration on September 8$^{\mathrm{th}}$ 2016 with 26.2 minutes of total time on-source.  The array configuration provided baselines ranging between respectively 15 and 463 meters, and between 15 and 2483 meters.  During the observations the precipitable water vapor had a median value at zenith of respectively 1.64 and 0.56~mm.  

    Two of the four spectral windows of the ALMA correlator were configured in Time Division Mode (TDM) to maximise the sensitivity for continuum observations (128 channels over 1.875~GHz usable bandwidth).  These two TDM spectral windows were centered at 234.16~GHz and 216.98~GHz.  The other two spectral windows were configured in Frequency Division Mode (FDM) to target the $^{12}$CO J=2-1, $^{13}$CO J=2-1 and C$^{18}$O J=2-1 lines with a spectral resolution of 61~kHz, 122~kHz, and~122 kHz respectively.  The data were calibrated and combined using the \textit{Common Astronomy Software Applications} pipeline \citep[CASA, ][version 4.7.2]{2007ASPC..376..127M}.  
    
    Inspection of the calibrated visibilities shows a 16\% difference in amplitude between the two observations at short baselines.  We assume that the emission from the midplane is constant in the 4.5 month period spanning the observations and that the difference is due to calibration uncertainties. 
    \gpp{Inspection of the calibrator archives does not lead us to favour one calibration over the other, and we decide to scale the flux of the compact array configuration to match the extended array configuration data.}    We estimate the absolute flux calibration to be accurate within $\sim$20\%, details of the observations and calibration are summarized in Table \ref{table_obs}.

\begin{table*}
\caption{Details of the observations.\label{table_obs}} \smallskip
\begin{minipage}[t]{\textwidth}
\centering
\noindent\begin{tabularx}{\columnwidth}{@{\extracolsep{\stretch{1}}}*{7}{l}@{}}
\hline\hline
UT Date & Number & Baseline Range & pwv &  \multicolumn{3}{c}{Calibrators:} \\ 
 & Antennas & (m) & (mm) & Flux & Bandpass & Gain  \\
\hline
2016 Apr 23	& 42 & 15 to 463  & 1.64 &  J1107-4449 &  J1107-4449 &  J1132-5606\\ 
2016 Sep 08	& 36 & 15 to 2483 & 0.56 &  J1107-4449 &  J1107-4449 &  J1132-5606 \\
\hline
\end{tabularx}
\end{minipage}
\end{table*}

\begin{figure}
       	\centering
        \includegraphics[width=\hsize]{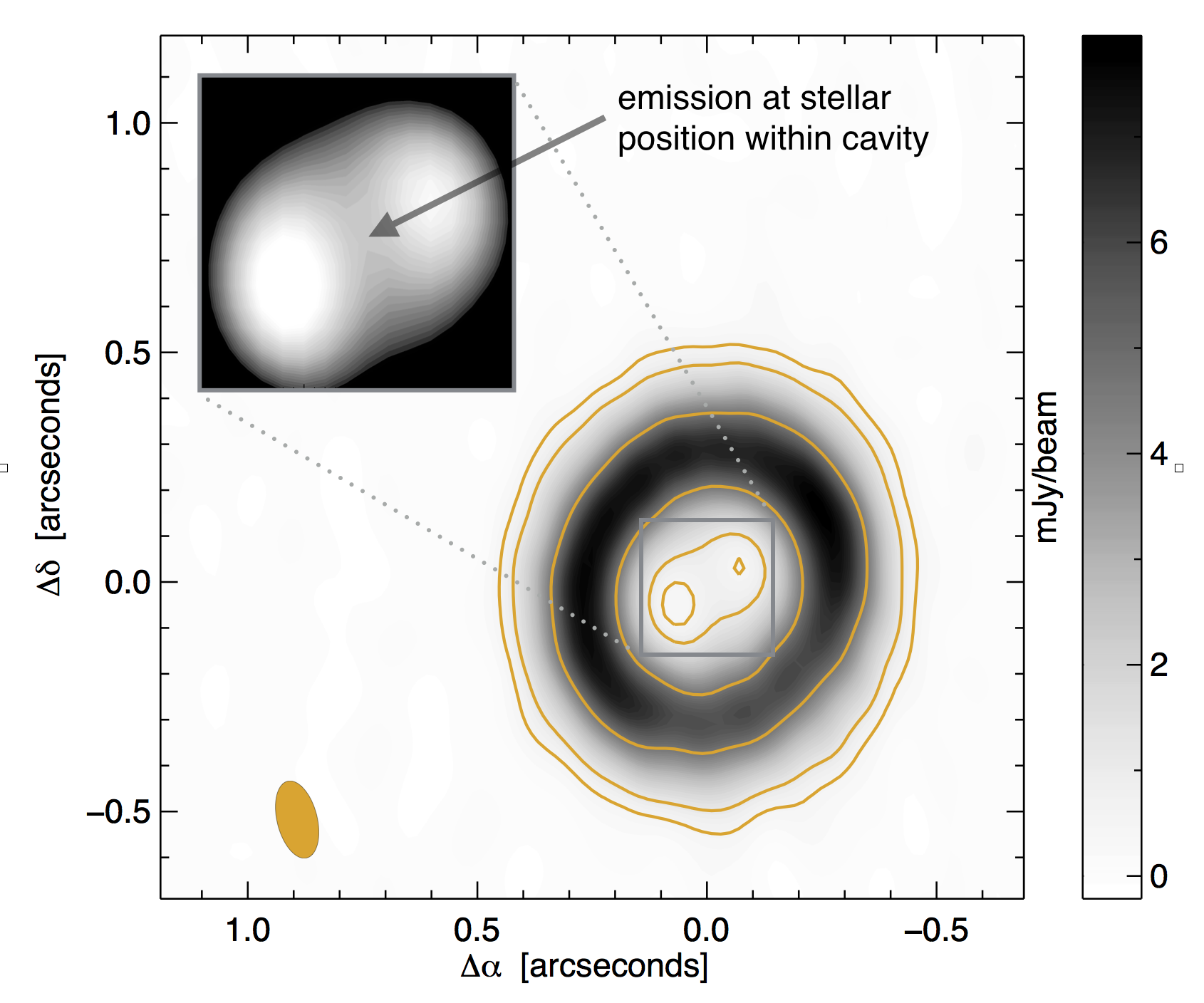}
        \caption{Continuum image of \object{HD~100453} for the ALMA band 6 observations, reconstructed using superuniform weighting resulting in a 0\farcs09 x 0\farcs17 beam.  Over plotted are contours at 12, 25 and 100 times the RMS value of 0.05 mJy/beam.  The beam is shown in orange in the bottom left, and a 0\farcs3 wide inset of the disk cavity with stretched colours highlights the emission at the stellar position. Note that the color scale is negative.}
        \label{fig:continuum}
    \end{figure}

    We imaged the continuum visibilities with the CLEAN task in CASA \citep{1974AAS...15..417H} using Briggs and superuniform weightings, which results in a restored beam size of respectively  0\farcs23 $\times$ 0\farcs15 at PA = 25.1 degrees (Briggs) and 0\farcs17 $\times$ 0\farcs09 at PA = 14.1 degrees (superuniform).  The dynamic range of these images is limited by the bright continuum source and \gp{we performed two rounds of phase only} self-calibration, resulting in a final RMS of 0.05 mJy/beam \gp{(peak SNR ratio of 159)} for the images created using superuniform weighting, and 0.04 mJy/beam \gp{(peak SNR ratio of 362)} for the images created using Briggs weighting.  We show the resulting continuum map in Figure \ref{fig:continuum}.  
	
\begin{figure*}
        \centering
        \includegraphics[width=\hsize]{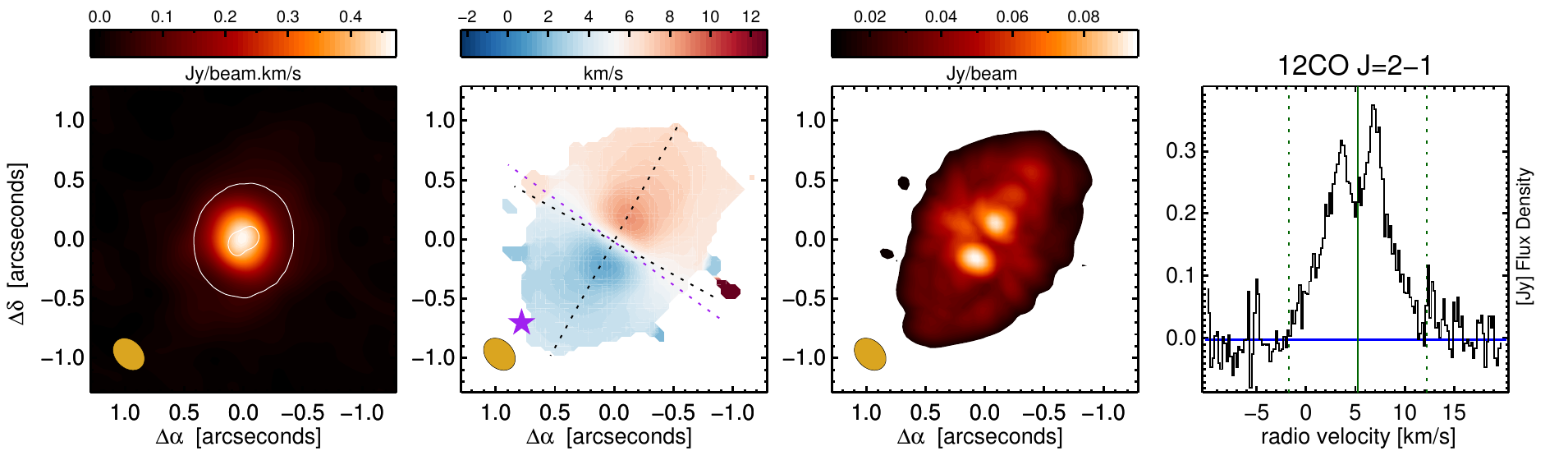}
        \caption{Summary of the \element[ ][12]{CO} line emission in \object{HD 100453}.  We show the integrated intensity (moment 0, left panel), intensity-weighted velocity (moment 1, 2$^{\mathrm{nd}}$ panel), peak intensity (moment 8, 3$^{\mathrm{rd}}$ panel) and the integrated emission line (right panel).  The moment 1 + 8 maps were made using a 3 $\sigma$ cutoff from images reconstructed using natural weighting to maximize sensitivity.  Over plotted in the 1$^{\mathrm{st}}$ panel is the 25 $\sigma$ (1.27 mJy/beam) contour of the continuum emission shown in Figure \ref{fig:continuum}.  The beam is shown in orange in the bottom left of each panel.  We show the approximate position of HD~100453~B \gp{during our observations \citep[1\farcs05 at PA = 132\degr][]{2018ApJ...854..130W}} with a purple star in the 2$^{\mathrm{nd}}$ panel, together with two dotted lines that show the major and minor disk axis of a disk with a semi major axis value listed in Table \ref{tab:flux_lines} and the inclination and position angle determined from fitting the continuum emission.  \gp{The purple line highlights the clockwise rotation of the velocity field discussed in Section \ref{sec:rt}}. The line profile shown in the right panel shows the integration boundaries used to calculate the total line emission (a half line width of 7.0 km s$^{-1}$), the systemic velocity of 5.25 km s$^{-1}$, and the level of the continuum emission used to calculate the integrated line flux.}
        \label{fig:moments-12co21-100453}
    \end{figure*}

\begin{figure*}
        \centering
        \includegraphics[width=\hsize]{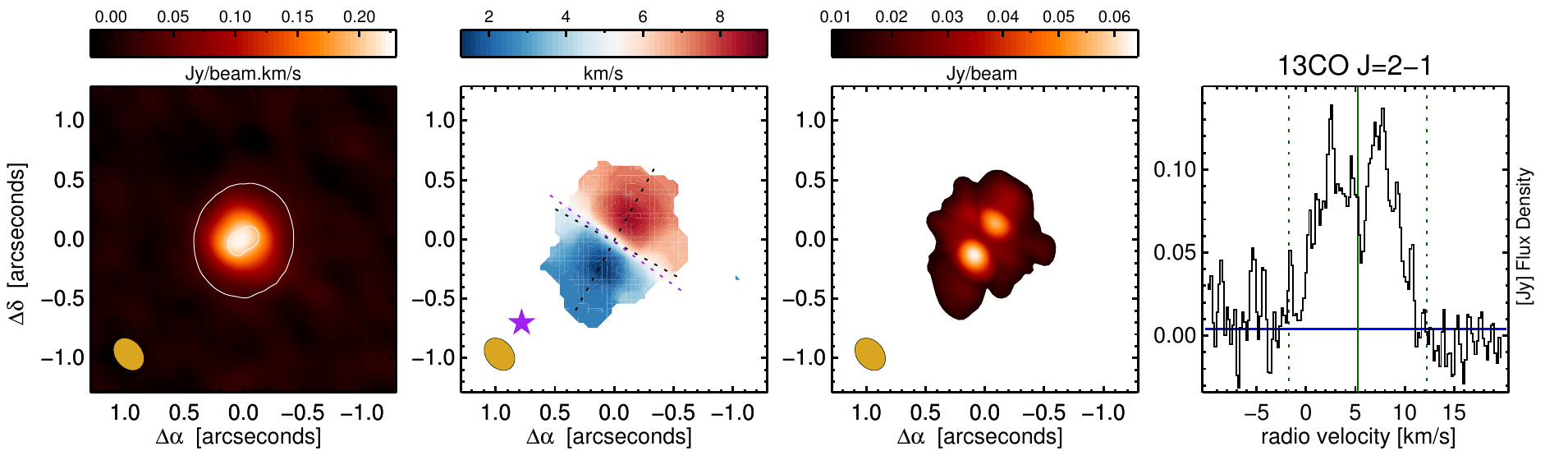}
       	\caption{Same as Figure \ref{fig:moments-12co21-100453} but for the \element[ ][13]{CO} J=2-1 line emission.}
        \label{fig:moments-13co21-100453}
    \end{figure*}

\begin{figure*}
        \centering
        \includegraphics[width=\hsize]{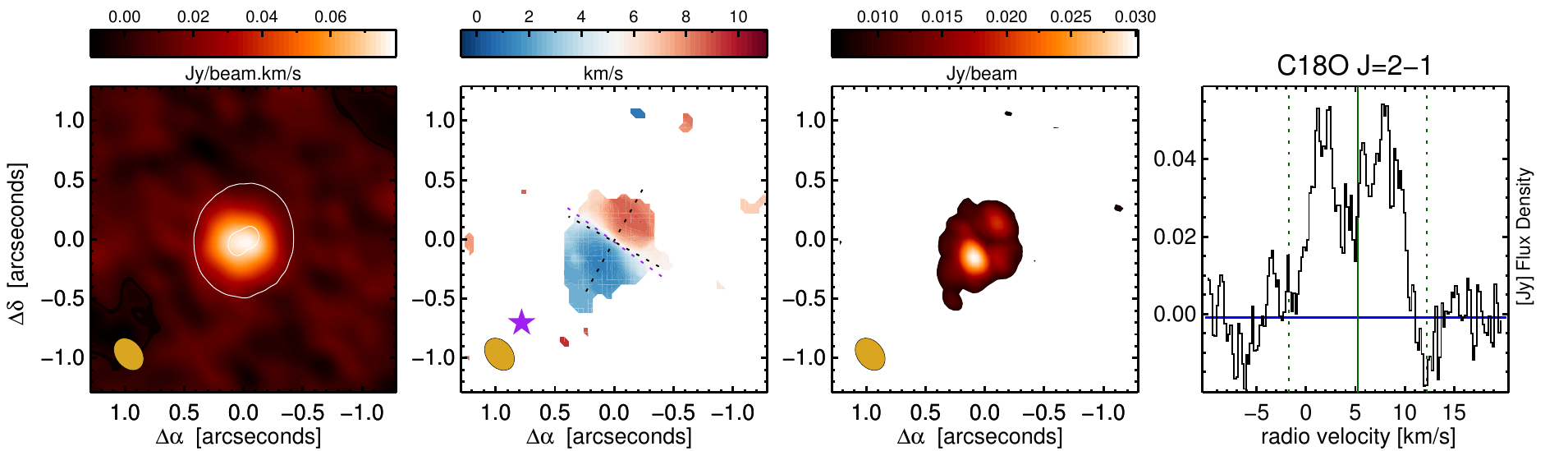}
        \caption{Same as Figure \ref{fig:moments-12co21-100453} but for the C\element[ ][18]{O} J=2-1 line emission.}
        \label{fig:moments-c18o21-100453}
	\end{figure*}

	We applied the self-calibration solutions obtained for the continuum emission to the CO visibilities and subtracted the continuum emission using the CASA task \textit{uvcontsub}.  We image the line data with a velocity resolution of 0.2 km s$^{-1}$ using natural weighting to maximise sensitivity which results in a restored beam of   0\farcs29 $\times$ 0\farcs23.  The CO line emission detections are summarised using the integrated intensity (moment 0), intensity-weighted mean velocity (moment 1) and peak intensity (moment 8) maps as well as the integrated spectra. These are shown in Figures \ref{fig:moments-12co21-100453}, \ref{fig:moments-13co21-100453} and \ref{fig:moments-c18o21-100453} for the \element[ ][12]{CO}, \element[ ][13]{CO} and C\element[ ][18]{O} J=2-1 transitions, respectively.

\section{Results}\label{sec:res}

    We detect and resolve the 1.4 mm dust continuum emission and the \element[ ][12]{CO}, \element[ ][13]{CO} and C\element[ ][18]{O} J=2-1 emission lines.  We determine the continuum flux and geometry by fitting several disk components to the visibilities and report the measured fluxes and derived geometry for the both the dust and gas emission in this section. We also report upper limits for the emissions coming from the location of HD~100453~B.

	\subsection{1.4 mm continuum emission}\label{res:continuum}
	
	    The dust continuum emission of HD~100453 shown in Figure \ref{fig:continuum} is concentrated into a ring that peaks at 0\farcs32 with \gp{an azimuthal variation along the ring of $\approx$ 30\% between the maximum at PA = 331\degr~and the minimum at PA = 180\degr}.  There is also excess emission present at the stellar position that, when convolved with the beam, connects with the outer disk along the beam major axis (see the inset in Figure \ref{fig:continuum}).
	    
	    \subsubsection{Disk geometry}\label{sec:geometry}
	
	        \gp{We use the fitting library \textit{uvmultifit} \citep{2014A&A...563A.136M} to quantify the inclination, position angle and spatial distribution of the disk emission. From our first look it is apparent that the emission can be broken up into several components, and we start by fitting the most obvious component (i.e., a uniform disk) to the visibilities after which we progressively add components to the model based on the imaged residuals. We end up using the following components to reach a satisfactory fit (i.e. no more recognisable structure in the residual emission): [1] a disk with a uniform surface brightness, [2] a ring, [3] a Gaussian, and [4] a point source. The order in which these components are added does not influence the final fitting results. }	  	     
	        
	        The components for the disk, ring and central component all share the same offset in RA and DEC from the phase center, the axis ratio, and the position angle, while the flux and semi major axis are left unconstrained.  We fit these geometries for each of the continuum windows (at 217~GHz and 234~GHz) separately to allow the detection of possible changes in flux due to the spectral slope of the dust emission $\alpha$ (S$_\nu$ $\propto$ $\nu^{-\alpha}$).
	
\begin{figure*}
                \centering
                \includegraphics[width=\hsize]{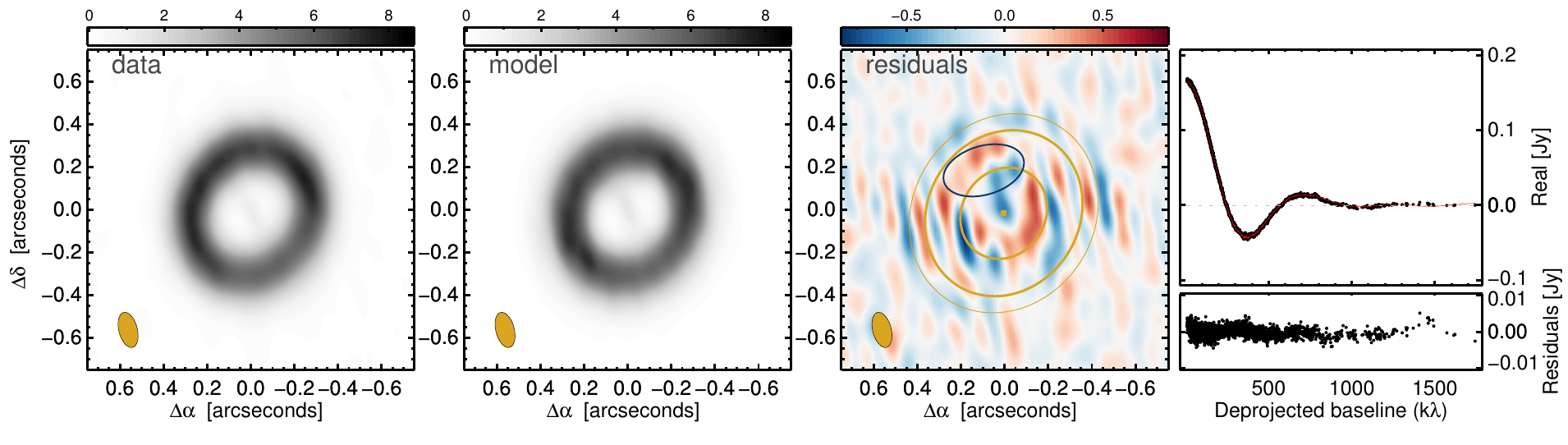}
                \caption{Comparison of ALMA band 6 data (left panel) with the best-fit composite model (2$^{\mathrm{nd}}$ panel).  The 3$^{\mathrm{rd}}$ panel shows the imaged residual visibilities.  This panel also includes ellipses representing the fitted disk and the central components in yellow thick lines, the outer ring with a yellow thin line, and the Gaussian component with a dark solid line.  Units of all intensity scales are in mJy/beam. The top right panel shows the real part of the visibilities as function of the deprojected baseline for the data (black dots) and the model (red line).  The bottom panel shows the residuals.  The visibilities are binned in sets of 200.} 
                \label{fig:model_comparison}
            \end{figure*}

	        We achieve the best fit with a combination of a disk, ring, Gaussian and a point source component (Figure \ref{fig:model_comparison}).  The Gaussian component is offset from the center of the cavity with 0\farcs09 and 0\farcs20 in RA and DEC respectively, and has a semi major axis of 0\farcs19, an axis ratio of 0.61, and a PA of 104.7\degr.  This feature overlaps with the northern spiral arm detected in scattered light and we discuss it further in the next Section (\ref{res:spiral}).
	
    	    The flux of the unresolved central component at 234~GHz is 1.3 $\pm$ 0.1 mJy, the combined flux of all components is 149.2 $\pm$ 3.0 mJy.  The uniform disk is constrained between 0\farcs22 and 0\farcs40 and inclined by 29.5 $\pm$ 0.5\degr with a position angle of 151.0 $\pm$ 0.5\degr.  An unresolved ring of emission at 0\farcs48 $\pm$ 0\farcs01 containing $\approx$ 13\% of the total flux improves the fit to the visibilities further.  It is unclear from our data whether this represents a real structure such as a second ring or spiral arms, or that it is an artefact of our use of a uniform disk with a discontinuity in flux at the inner and outer edge (i.e. the unresolved ring takes the place of a tapered or power-law outer disk).  The spectral index for the disk component is 2.4 $\pm$ 0.1, and between 3.0 and 3.6  for the other components, respectively. 

            The best fit parameters for the fitted components are summarized in Table \ref{tab:flux} and visualized in Figure \ref{fig:model_comparison}, where we compare the imaged model and residuals to the HD~100453 disk and show the real part of the visibilities for the data, the model, and their difference.

\begin{table*}
\caption{Best-fit parameters with their respective 1 $\sigma$ uncertainty in parenthesis, obtained from fitting components to the continuum visibilities: A disk with a radially constant surface brightness, an unresolved ring, a point source, and a Gaussian. The center, position angle and inclination for the 3 first components have been fixed during the fitting. The spectral slope $\alpha$ (5$^{\mathrm{th}}$ column) is calculated following S$_\nu$ $\propto$ $\nu^{-\alpha}$ using measurements at 234.2 and 217.0~GHz (1.28 and 1.38 mm). \label{tab:flux}} 
\begin{minipage}[t]{\textwidth}
\centering
\noindent\begin{tabularx}{\columnwidth}{@{\extracolsep{\stretch{1}}}*{8}{l}@{}}
\hline\hline
Component & $\Delta$RA & $\Delta$DEC & S$_{v, 234.2 GHz}$ & $\alpha$ & semi major axis & inclination & PA   \\ 
   & [\arcsec]    & [\arcsec]     & [mJy]  &  & [\arcsec]         & [\degr]       & [\degr]         \\
\hline
Disk     & -$^{\mathrm{a}}$ & -$^{\mathrm{a}}$ & 152.5 (0.5) & 2.4 (0.1) & 0.22 (0.01), 0.40 (0.01)$^{\mathrm{b}}$ & 29.5 (0.5) & 151.0 (0.5)   \\
Ring     & -$^{\mathrm{a}}$ & -$^{\mathrm{a}}$  & 18.9  (0.6)  & 3.0 (0.4) & 0.48 (0.01)      & fixed     & fixed        \\
Point    & -$^{\mathrm{a}}$ & -$^{\mathrm{a}}$  & 1.3  (0.1)  & 3.5 (1.3) & -     & -     & -       \\
Gaussian & 0.09$^{\mathrm{c}}$ & 0.20$^{\mathrm{c}}$ & 8.7 (0.3) & 3.6 (0.6) & 0.19 (0.01)     & 52.2 (5.0) & 104.7 (3.0)   \\
\hline
All components    &  &  &  149.2 (3.0) & 2.6 (0.1) &   &      &       \\ 
\hline
\end{tabularx}
\end{minipage}
\tablefoot{\textbf{$^{\mathrm{a}}$:} \gp{The first three components have been fixed to the center fitted for the disk component}. \textbf{$^{\mathrm{b}}$:} Contains two values for the disk component: the inner and outer radius. \textbf{$^{\mathrm{c}}$:} Offset relative to the center of the best-fit disk and ring component.}
\end{table*}

	    \subsubsection{A mm counterpart to the northern spiral arm or a vortex?}\label{res:spiral}
\begin{figure*}
                \centering
           	    \includegraphics[width=\hsize]{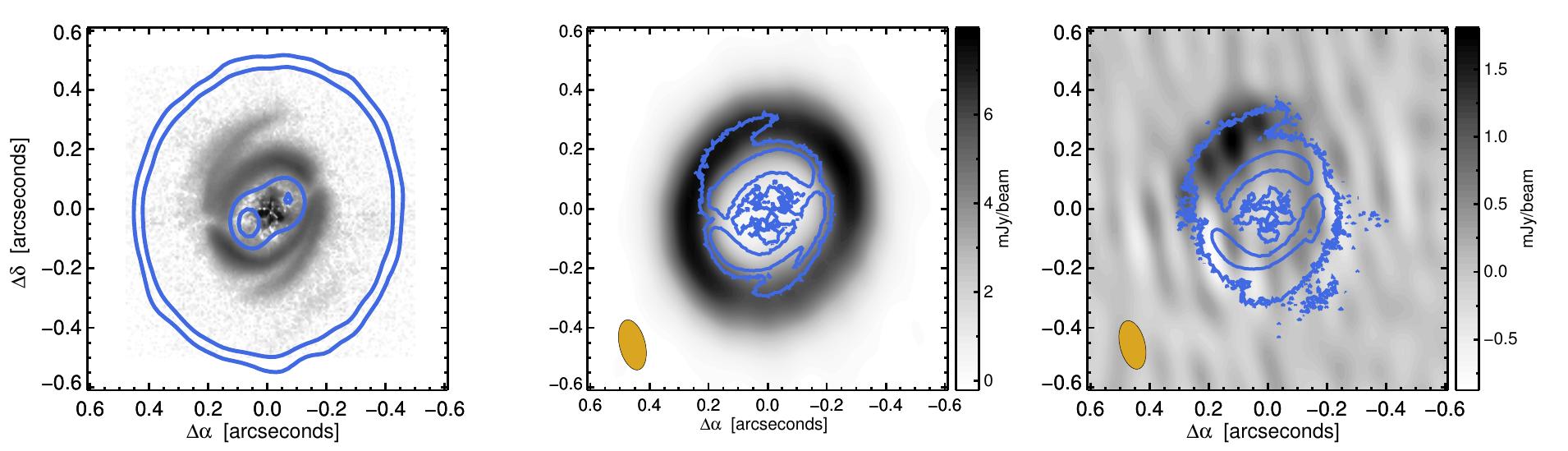}
                \caption{\textbf{Left panel:} J-band Q$_\phi$ image reproduced from \citet{2017A&A...597A..42B} in arbitrary intensity units with an overlay of the \gp{12 and 25 $\sigma$} contours of the ALMA data presented in Figure \ref{fig:continuum}.  \textbf{Central panel:} Inverted counterpart to the images shown in the left panel but with arbitrary contours of the SPHERE data overlayed on the ALMA data.  \textbf{Right panel:} ALMA residuals after subtracting the best-fit disk, ring and central point source components summarized in Table \ref{tab:flux} in visibility space, imaged using superuniform weighting.  The same contours as shown for the SPHERE images in the central panel are again overlayed. The Gaussian component of the ALMA continuum emission appears to coincide well with the Northern spiral arm seen in scattered light.}
           	    \label{fig:alma-sphere}
       	    \end{figure*}

	        There is significant residual emission at the same location as the northern spiral arm detected in scattered light when only considering axisymmetric components for the disk. These residuals can be fit with a single elliptical Gaussian containing 8.7 mJy of flux at the same position and with a similar positioning as the northern spiral arm as seen in scattered light (Table \ref{tab:flux} and the right panel of Figure \ref{fig:alma-sphere}). 
	        
	        \gp{To better compare this emission to the spirals detected in scattered light we subtract the best-fit disk, ring, and central component from the data in visibility space and image the residuals.  We show these residuals together with the SPHERE image published in \citet{2017A&A...597A..42B} in the right panel of Figure \ref{fig:alma-sphere}. The other two panels in that Figure show the two datasets imposed over each other to illustrate their relative spatial extent.

            All the scattered light emission including the two spiral arms is contained within the region where mm emission is detected, with the bulk of the scattered light emission originating from within the mm emission disk cavity.  Comparison of the cavity outer radius with the scattered light data presented by \citet{2017A&A...597A..42B}, as shown in the left panel of Figure \ref{fig:alma-sphere}, highlights a striking similarity between the two datasets in their deviation from circular symmetry. Both maps show an almost hexagonal shape of the cavity border suggesting that whatever mechanism is shaping the disk cavity is not acting in an azimuthally symmetric way.

            The mm residual emission is unresolved in the radial direction, recovered from our data regardless of the weighting applied during the imaging, and matches both the radial extent and positioning of the northern spiral arm (Figures \ref{fig:alma-sphere}, right panel). Given the quality of our data, however, it is not clear whether this really is a mm counterpart to that spiral arm. Another viable origin for this emission would be a vortex such as detected in the HD~135344B disk \citep{2016ApJ...832..178V}. That vortex is co-spatial with the end of the spiral arm detected in scattered light, and in itself is likely responsible for launching the spiral arm due to its mass, \gpp{while being induced by an interior body} \citep{2018A&A...619A.161C}. 
            
            Future higher resolution observations are needed to disentangle the nature of the excess mm emission. However, both scenarios mentioned above lead to the same conclusion which we will explore in the remainder of this manuscript: that HD~100453B does not induce the twin spiral arms seen in scattered light. Either the excess mm emission comes from a vortex which in itself induces the northern spiral arm, or it is the mm counterpart of the northern spiral arm. This latter option makes the northern spiral arm the primary arm (containing most mass) which is inconsistent with the position and orbital motion previously derived for of HD~100453B.  We only refer to the spiral arms scenario in the following analysis and discussion sections in order to keep them as concise as possible, and reiterate in our conclusions the two likely scenarios for the excess emission.}

	    \subsubsection{Dust mass estimates for the circum-primary and circum-secondary disks}\label{sec:dustmass}
	
	        To convert the measured continuum emission into a dust mass we assume that the emission is optically thin and of a single temperature following 
	
	        \begin{equation}\label{eq:dustmass}
            log M_{dust} = log S_{\nu} + 2 log d - log \kappa_{\nu} - log B_{\nu}( \langle T_{dust} \rangle),
            \end{equation}
        
            where $S_{\nu}$ is the flux density, $d$ is the distance, $\kappa_{\nu}$ is the dust opacity, and  $B_{\nu}(\langle T_{dust} \rangle)$ is the Planck function evaluated at the average dust temperature \citep{1983QJRAS..24..267H}.  We adopt a dust opacity of 2.31 $cm^{2}~g^{-1}$ at 1.28 mm, calculated using astronomical silicate \citep{1984ApJ...285...89D,1993ApJ...402..441L, 2000AAS...197.4207W}, with a grain size distribution with sizes between 0.1 and 3000 $\mu$m distributed following a power law with a slope of -3.5. 
            

            Typically the dust temperature is estimated extrapolating from the mass averaged dust temperature in grids of radiative transfer disk models that cover a range of stellar and disk parameters \citep[e.g.][]{2013ApJ...771..129A,2016ApJ...819..102V}.  At the moment these grids only consider "full disks" and thus do not give accurate dust temperatures for disks like the one around HD~100453~A which consists of a relatively narrow ring of dust.

            Instead we perform a radial decomposition of the disk intensity using the radiative transfer code \textsc{mcfost} \citep{2006A&A...459..797P,2009A&A...498..967P} as was previously done for HL~Tau by \citet{2016ApJ...816...25P}, to estimate the dust temperature in the disk. Shortly, we fix the disk inclination and PA and match the model radial surface density profile in an iterative procedure to the observed one. See Section 3 of \citet{2016ApJ...816...25P} for a full description. 
            The dust mass in the resulting model is 0.07  M$_{\mathrm{jup}}$ and the mass averaged dust temperature in the resulting model is 27~K which translates in a dust disk mass of 0.09 M$_{\mathrm{jup}}$ applying Equation \ref{eq:dustmass}.%

	        We do not detect any signal at the location of HD~100453~B.  We measure the continuum RMS in a circular region centered on the companion location with a diameter of 0\farcs20 (20.6~au) using the Briggs-weighted images for the best compromise between spatial resolution and sensitivity.  The RMS at the location of the companion is \gp{0.033 mJy, leading to a 3 $\sigma$ upper limit of 0.099}  mJy. 
	        To calculate a limit on the amount of dust that can be present around HD~100453~B we estimate the average dust temperature using the stellar luminosity determined from the BHAC2015 evolutionary tracks \citep{2015A&A...577A..42B} for a 0.2 M$_\odot$, 10 Myr old star.  The expected average dust temperature in a disk with an outer radius of 10 au around such a star is 22~K following Figure 5 of \citet{2016ApJ...819..102V}.  This puts an upper limit of 0.03 M$_{Earth}$ on the amount of dust around HD~100453~B.

	\subsection{CO J=2-1 isotopologue emission lines}\label{res:gas} 

	    We detect spatially and spectrally resolved emission from the \element[ ][12]{CO}, \element[ ][13]{CO} and C\element[ ][18]{O} J=2-1 emission lines from the HD~100453 disk and show the moment maps and line profiles in Figures \ref{fig:moments-12co21-100453}, \ref{fig:moments-13co21-100453} and \ref{fig:moments-c18o21-100453}, respectively.
	
	    We estimate the systemic velocity from the \element[ ][12]{CO} J=2-1 emission line at v$_{LSR}$ = 5.25 $\pm$ 0.10 km s$^{-1}$, based on the center of the line profile and the channel maps.  The  \element[ ][12]{CO} emission line is detectable up to projected velocities of $\pm$~7.0 km s$^{-1}$ from the systemic velocity, which translates to a distance from the central star of 7.4 au assuming the gas is in Keplerian rotation in a disk inclined by 29.5\degr around a 1.70 M$_{\odot}$ star\footnote{\gp{This is an upper limit as both beam dilution as higher velocity gradients make the CO emission more difficult to detect at higher velocities and at closer distance.}}.  The outer radius as measured from \element[ ][12]{CO} emission above 3$\sigma$ in the moment maps is 1\farcs10.  We make a first order estimate of the disk inclination using the axis ratio measured from the moment maps, and find that the inclination for the \element[ ][13]{CO} (31 $\pm$ 5\degr) and C\element[ ][18]{O} (35 $\pm$ 5\degr) emission is in agreement with the inclination determined from the continuum data, while the \element[ ][12]{CO} emission appears more inclined (49 $\pm$ 5\degr).  The ratio of the line flux from the \element[ ][12]{CO}, \element[ ][13]{CO} and C\element[ ][18]{O} J=2-1 emission lines is 6.0:2.3:1.0 which is similar to the isotopologue ratio detected from more massive disks around other Herbig Ae/Be stars \citep[e.g.][]{2015ApJ...798...85P} and indicates that the CO emission is optically thick for at least the \element[ ][12]{CO} emission. We make an estimate of the optical depths for each isotopologue from the detected line ratios under the assumption that the emission comes from an isothermal slab \citep[see for details e.g. section 3.3 in][]{2015ApJ...798...85P}. We adopt a \element[ ][12]{CO} to \element[ ][13]{CO} ratio of 76 \citep{2008A&A...477..865S} and a \element[ ][12]{CO} to C\element[ ][18]{CO} ratio of 500 \citep{1994ARA&A..32..191W} and find optical depths of $\tau_{12CO}$ $\approx$ 39, $\tau_{13CO}$ $\approx$ 0.5 and $\tau_{C18O}$ $\approx$ 0.1.  The integrated line fluxes, spatial extent and geometry of all CO line emission are summarized in Table \ref{tab:flux_lines}. 


\begin{table*}
\setlength{\tabcolsep}{4pt} 

\caption{Line fluxes, spectral resolution, spatial extent and inclination for the CO J=2-1 isotopologue emission\label{tab:flux_lines}} 
\smallskip
\centering   
\begin{minipage}[t]{\textwidth}
\noindent\begin{tabularx}{\columnwidth}{@{\extracolsep{\stretch{1}}}*{7}{l}@{}}
\hline

Line  & line flux & error$^{\mathrm{a}}$ & channel width          & RMS$^{\mathrm{b}}$   & radius $^{\mathrm{c}}$ & \textit{i}$^{\mathrm{c}}$ \\
      & Jy  km s$^{-1}$  &  Jy   km s$^{-1}$ & m s$^{-1}$  & mJy/beam          &  \arcsec & \degr \\
\hline

$^{12}$CO J = 2-1      &  2.34 & 0.06 & 200 & 3.0 & 1.10 & 49 $\pm$ 5 \\
$^{13}$CO J = 2-1      &  0.90 & 0.03 & 200 & 3.0 & 0.70 & 31 $\pm$ 5 \\
C$^{18}$O J = 2-1      &  0.39 & 0.02 & 200 & 2.1 & 0.50 & 35 $\pm$ 5 \\

\hline                                   
\end{tabularx}
\end{minipage}
\tablefoot{Line fluxes have been calculated from the natural-weighted images by integrating the emission around the  systemic velocity at 5.25 km s$^{-1}$ assuming a half line width of 7.0 km s$^{-1}$. \textbf{$^{\mathrm{a}}$:} The error on the integrated line flux was estimated from the RMS of the integrated spectrum outside the line boundaries and does not include calibration uncertainties. \textbf{$^{\mathrm{b}}$:} 1 $\sigma$ RMS per channel. \textbf{$^{\mathrm{c}}$:} The radius is measured along the semi-major axis of the moment 1 maps shown in the 2$^{nd}$ panel of Figs. \ref{fig:moments-12co21-100453} - \ref{fig:moments-c18o21-100453} that were made using CO emission detected above 3$\sigma$ in the channel maps.} 
\end{table*}	    
	    
	    The velocity field of the disk is globally coherent with keplerian rotation although there are hints of a deviation present in the outer disk \gp{where the CO emission at systemic velocity appears to be rotated clockwise by several degrees.  To highlight this rotation of the velocity field we show the disk major and minor axis as determined from the dust emission together with a line following approximately the emission at zero projected velocity to guide the eye in the second panel of Figures \ref{fig:moments-12co21-100453} to \ref{fig:moments-c18o21-100453}.}  Furthermore, despite the presence of a small misaligned inner disk the velocity map of the CO emission lacks the typical 's' shaped pattern expected at the location of the warped inner disk as described by e.g. \citet{2014ApJ...782...62R} and detected in other Herbig Ae/Be disks such as HD~142527 \citep{2015ApJ...811...92C} and HD~97048 \citep{2017A&A...597A..32V}. The fact that the velocity field inside the cavity appears to be consistent with Keplerian rotation despite the presence of a misaligned inner disk in the cavity is possibly due to insufficient spatial resolution or to a lack of sensitivity of our observations. We explore possible deviations from Keplerian rotation in the disk further in Sect.  \ref{sec:rt}.

        As already remarked upon by \citet{2018ApJ...854..130W} the disk rotation direction is counter-clockwise if we follow the interpretation by \cite{2017A&A...597A..42B} that the faint spiral structure seen towards the SW of the disk in scattered light is actually scattering from a spiral arm on the opposite face of the disk and thus that the SW part of the disk is the side nearest to us.  This means that the spiral arms seen in scattered light are trailing.

        \subsubsection{CO gas mass and the gas to dust ratio}\label{sec:discussion_gtd}
        
            Deriving a total gas mass from CO emission is a highly uncertain endeavour given the large uncertainties on, among other things, the local conditions at the emitting surface, the amount of CO in gas phase, and the conversion between CO mass and total gas mass \citep[see e.g.][]{2017A&A...599A.113M,2018ApJ...864...78K}. Large parametric disk grids relating a suit of disk parameters to simulated observable CO line fluxes can somewhat alleviate these uncertainties. We use the grid of disk models calculated by \citet{2014ApJ...788...59W} to estimate a total gas mass based on the isotopologue CO line ratio for the disk around HD~100453~A between 0.001 and 0.003 M$_\odot$ depending on the \element[ ][12]{CO}/C\element[ ][18]{O} ratio assumed (550 or 1650). Using the dust mass of 0.07 M$_{\mathrm{jup}}$ derived in Section \ref{sec:dustmass} we arrive at a gas to dust ratio of 15 to 45. 
        
            This value is in agreement with a previous upper limit on the disk gas mass by \citet{2009ApJ...697..557C} who suggested that the outer disk is significantly depleted in gas with an estimated gas to dust ratio between a few and a few tens.%
\section{Analysis}\label{sec:ana}

	One of the reasons why the HD~100453 system is of interest is the possible connection between the two spiral arms detected in scattered light and the 0.2 M$_{\odot}$ companion orbiting at a projected distance of 1\farcs05 (108 au) from the central star.  Such a companion, if in a low eccentricity and close to co-planar orbit, would excite 2 spiral arms similar to those detected in scattered light \citep{2016ApJ...816L..12D, 2018ApJ...854..130W}.  This tidal interaction would also truncate the circumprimary (CP) disk at a fraction of between $\approx$ 1/2 and 1/3 of the semi-major axis \citep{1994ApJ...421..651A}, in agreement with the outer radius of the disk as detected in scattered light and mm continuum emission.
	
	However, our observations bring several discrepancies with this interpretation. The \element[ ][12]{CO} gas disk extends to 1\farcs10 (113 au) and overlaps with the projected position of the secondary. 
	Furthermore, hydro simulations for spiral arms induced by co-planar orbiting planets indicate that the surface density enhancement is expected to be higher in the primary arm, i.e., the one pointing to the perturber, than in the secondary arm \citep{2015ApJ...815L..21F}, which means the southern spiral arm is the primary if it were induced by a co-planar companion. Yet, we only detect mm emission from the location of northern spiral arm (c.f. Section \ref{res:spiral}). If the mm continuum excess detected in the Northern arm is indicating that this arm is the more massive one then, because it is not pointing to the perturber, it is not clear anymore that the spirals are driven by the companion M star, in particular if it is in a co-planar and prograde orbit with the disk.  
	
	Lastly, despite the proximity of the companion to the CP disk we detect no emission from a circumsecondary or circumbinary disk. This is contrary to the idea that a recent flyby, prograde and co-planar, \gpp{as such an interaction} would likely lead to a significant amount of dust and gas being captured by the by the interloper \citep{2018MNRAS.tmp.3163C}.\\  

	If any of the three arguments above is correct, it would challenge the proposed co-planarity of the orbit  of the companion and its dominant role in the excitation of the spiral arms.  We investigate the influence of a co-planar orbit companion using gas+dust SPH simulations in Section \ref{sec:sph} and the possible deviation of the gas kinematics in the outer disk in Section \ref{sec:rt}.  We also re-assess the orbital parameters to further check the viability of a co-planar orbit for the HD~100453~AB system in Section \ref{dis:orbit}.
	
	\subsection{SPH Simulations}\label{sec:sph}
	
	    We study the tidal influence of a co-planar companion on the gas and dust content of the circumprimary disk via global 3D simulations with the Smoothed Particle Hydrodynamics (SPH) code \textsc{Phantom} \citep{2018PASA...35...31P}. Gas and dust are treated as separate sets of particles interacting via aerodynamic drag according to the algorithm described in \citet{2012MNRAS.420.2345L}, using $7.5\times10^5$ SPH particles for the gas and $2.5\times10^5$ for the dust and setting the initial dust-to-gas mass ratio to 1\%. The grain size is set to 1~mm. We adopt for our simulations the same parameters for the binary orbit and for the disk as in \citet{2016ApJ...816L..12D}. The primary and secondary stars, treated as sink particles, have masses $M_\mathrm{A}=1.7$ and $M_\mathrm{B}=0.3$~M$_\odot$\footnote{Note that in this Section we follow \citet{2016ApJ...816L..12D} in using a companion mass of 0.3~M$_\odot$ \citep[from][]{2006A&A...445..331C}, whereas in the rest of this manuscript we adopted a companion mass of 0.2~M$_\odot$ \citep{2009ApJ...697..557C}. The impact of using a lighter companion in our simulations would be to decrease the size of its Roche lobe and the amount of mass the secondary can capture} and are separated by $a=120$~au on a circular orbit, co-planar with the disk. We initially set the inner and outer disk radii to $r_\mathrm{in}=12$ and $r_\mathrm{out}=96$~au (note that $r_\mathrm{out}$ is outside the Roche lobe of the primary) and its mass to $M_\mathrm{d}=0.003$~M$_\odot$, with power-law profiles $\Sigma\propto r^{-1}$ for the surface density and $T\propto r^{-0.5}$ for the temperature. Contrary to \citet{2016ApJ...816L..12D}, we do not seek here to reproduce the pitch angle of the spirals and adopt a more conventional disk aspect ratio of $H/r=0.05$ at 12~au , with a vertically isothermal profile. We set the SPH artificial viscosity in order to obtain an average \citet{1973A&A....24..337S} viscosity of $\alpha_\mathrm{SS}=5\times10^{-3}$ \citep{2010MNRAS.405.1212L}. The accretion radius of both stars is set to 12 and 5~au, respectively. We run the simulation for 10 orbits of the binary, at which time the disk has reached a quasi-steady state.
			
\begin{figure*}
            \centering
            \includegraphics[width=\hsize]{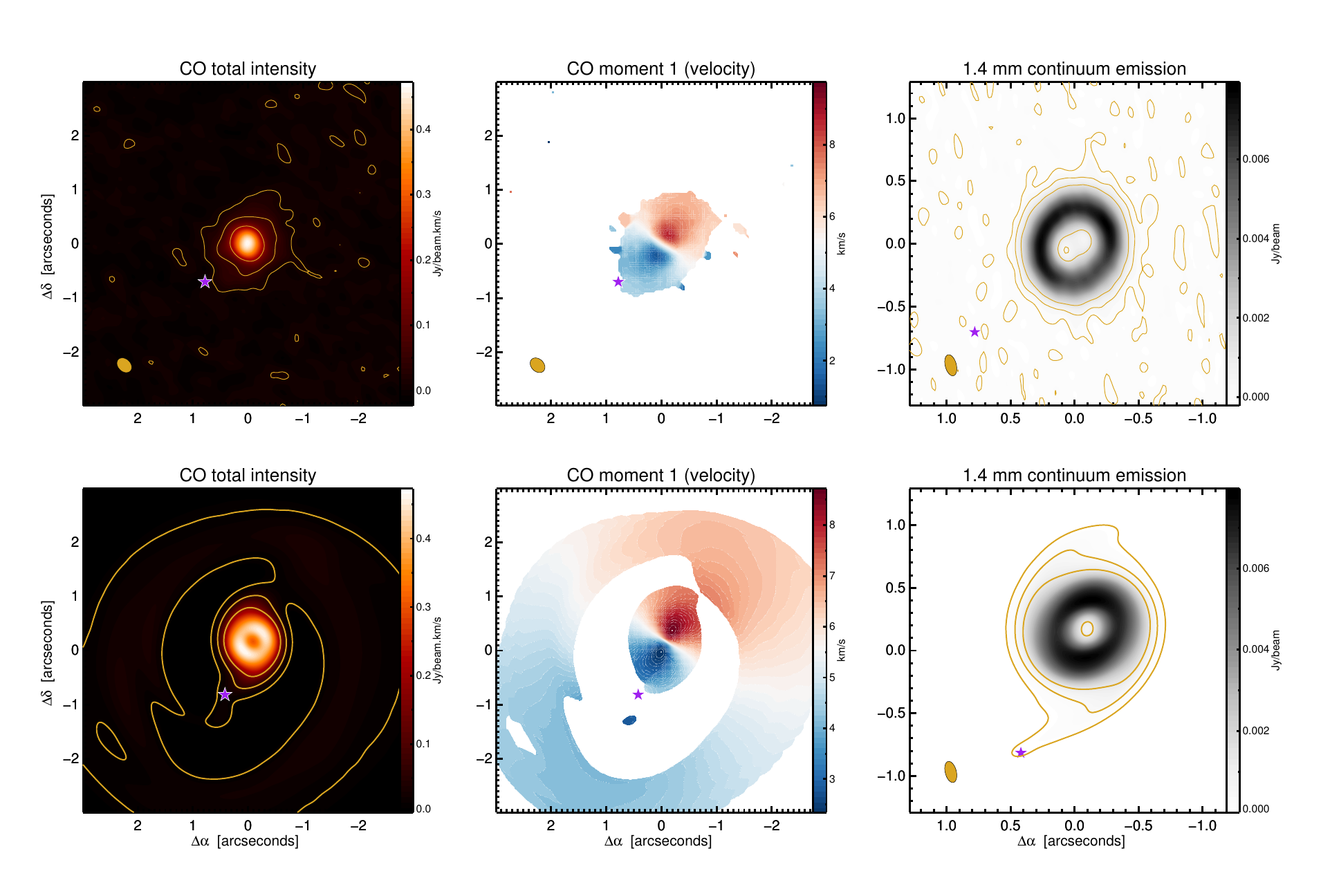}
            \caption{Comparison between the ALMA observations (top row) and the ray-traced SPH simulations (bottom row).  Panels from left to right: CO integrated intensity (moment 0) map, CO intensity weighted velocity field (moment 1) map, and the 1.4 mm dust emission map. The purple star represents the location of HD~100453~B in all panels.  For the CO moment 1 map (middle panel) we only include emission that is within a certain fraction of the peak emission in the image channels.  The maximum observed dynamic range in our observations is 40, and we construct the model moment 1 maps using only emission that is brighter than a fraction of 3/40 of the peak intensity.  In the top left and right panels we show 2, 10 and 30 sigma contours using yellow lines. For the bottom panels we use the dynamic range from the observations to approximate these contours as fraction of the maximum emission in the simulated maps.}
            \label{fig:sph_sim1}
        \end{figure*}

	    To facilitate a more quantitative comparison between our simulations and the observed data we use the radiative transfer code \textsc{mcfost} to convert the results of our simulation into images at relevant frequencies.  We convolve the resulting images with a Gaussian of the same FWHM as the beam in the observations and scale the maximum intensity in the convolved image to that of the observed images.  We show the simulated 1.4 mm continuum map, the integrated CO intensity, and the CO velocity field, together with the observed maps, in Figure \ref{fig:sph_sim1}. \\

	    In our simulations two other disks quickly form from the material that was part of the circumprimary disk and located outside the primary's Roche lobe: a circumsecondary and a circumbinary disk. In the simulated 1.4 mm map most of the continuum emission originates from the circumprimary disk. Compared to this disk the peak flux from the circumsecondary and circumbinary disks are weaker by a factor 80 and 300, respectively. For comparison, the observed ratio between the peak flux from the circumprimary disk and the background RMS is $\approx$ 400.  The dust grains in the circumprimary disk get concentrated in a smaller disk with two faint spiral arms whose primary arm is marginally brighter than the secondary arm. 	    The CO velocity field in the simulations shows a twist in the same manner as seen in the observations, \gpp{ and the CO disk becomes more elongated} as it fills the Roche lobe of the primary. We discuss these results further in Section \ref{res:synth}.

	\subsection{Quantifying the disk warp}\label{sec:rt} 

        The velocity field of the disk around HD~100453~A shows deviations from a pure Keplerian rotation, most notably through a twist in the iso-velocity contours at systemic velocity \gp{(highlighted in Figs. \ref{fig:moments-12co21-100453} to \ref{fig:moments-c18o21-100453} with a purple line)}.  To better quantify these deviations we fit the observed velocity field of the CO gas using the methodology introduced in \citet{2015ApJ...798...85P} which we shortly summarize in the next paragraph. We note that we restrict our analysis to quantifying the velocity field and the warp in the circumprimary disk. We do not optimize on the disk structure \gp{other than obtaining a reasonable fit} and will explore the intra-cavity column density and kinematics in an upcoming paper using higher sensitivity and resolution observations.\\
    	
        We fit a parametric model of the $^{12}$CO gas allowing for a warp (i.e., a different inclination) and PA \gp{of the inner disk} w.r.t the outer disk \gp{which starts} at 38 au.  The parametric model follows \citet{2015ApJ...811...92C} and adopts the surface density parameters fitted by \citet{2018ApJ...854..130W}  \gp{to the lower resolution compact array configuration part of the dataset also presented in this manuscript, with exception of the CO scale height ($H/r$), the power law for the radial surface density ($\gamma$), and the characteristic radius (r$_{\mathrm{c}}$). Following our choice for $H/r$ described in Section \ref{sec:sph} we choose a more conventional value for the disk aspect ratio of 0.05.  Because no value for $\gamma$ is mentioned in \citet{2018ApJ...854..130W} we use a standard value of 1. Finally, we are unable to reproduce the outer disk extent with a large value \gpp{for R$_{\mathrm{c}}$} of 27 $\pm$ 1 au, and instead use a value of 10 au which better reproduces the extent of the outer disk.}

        Our four free parameters are the inclination angle and PA for the inner and outer disk: $\{i_{\rm out}, {\rm PA}_{\rm out}, i_{\rm in}, {\rm PA}_{\rm in}\}$.  We compare the model and data via the computation of first moment maps. Optimization is done by minimizing $\chi^2 = \sum({\rm M1}^{\rm o}-{\rm M1}^{\rm m} )$, where ${\rm M1}^{\rm o}$ and ${\rm M1}^{\rm m}$ correspond to observed and model first moment. The comparison is done only in the pixels where the observed signal in the zeroth moment is above $5\sigma$.  First, we performed a simple $\chi^2$ search using 0.5~degrees steps around $\{i_{\rm out}, {\rm PA}_{\rm out}\} = {25, 140}$. We fix the outer disk values to those that yield a minimum in $\chi^2$. Then, we do the same search but for the intra-cavity angles $\{i_{\rm in}, {\rm PA}_{\rm in}\}$.  We adopt the best fit values and repeat the exploration for the outer disk parameters. We repeat the same for the intra-cavity angles.  These steps are repeated until the variation is $<0.5$ degrees (our step). The final best fit parameters are $\{i_{\rm out}, {\rm PA}_{\rm out}\} = \{19.5, 139.5\}$ for the outer disk parameters, and $\{i_{\rm in}, {\rm PA}_{\rm in}\} = \{24.0, 146.0\}$ \gpp{ for the inner disk}. \\
        
\begin{figure*}
            \centering
            \includegraphics[width=\hsize]{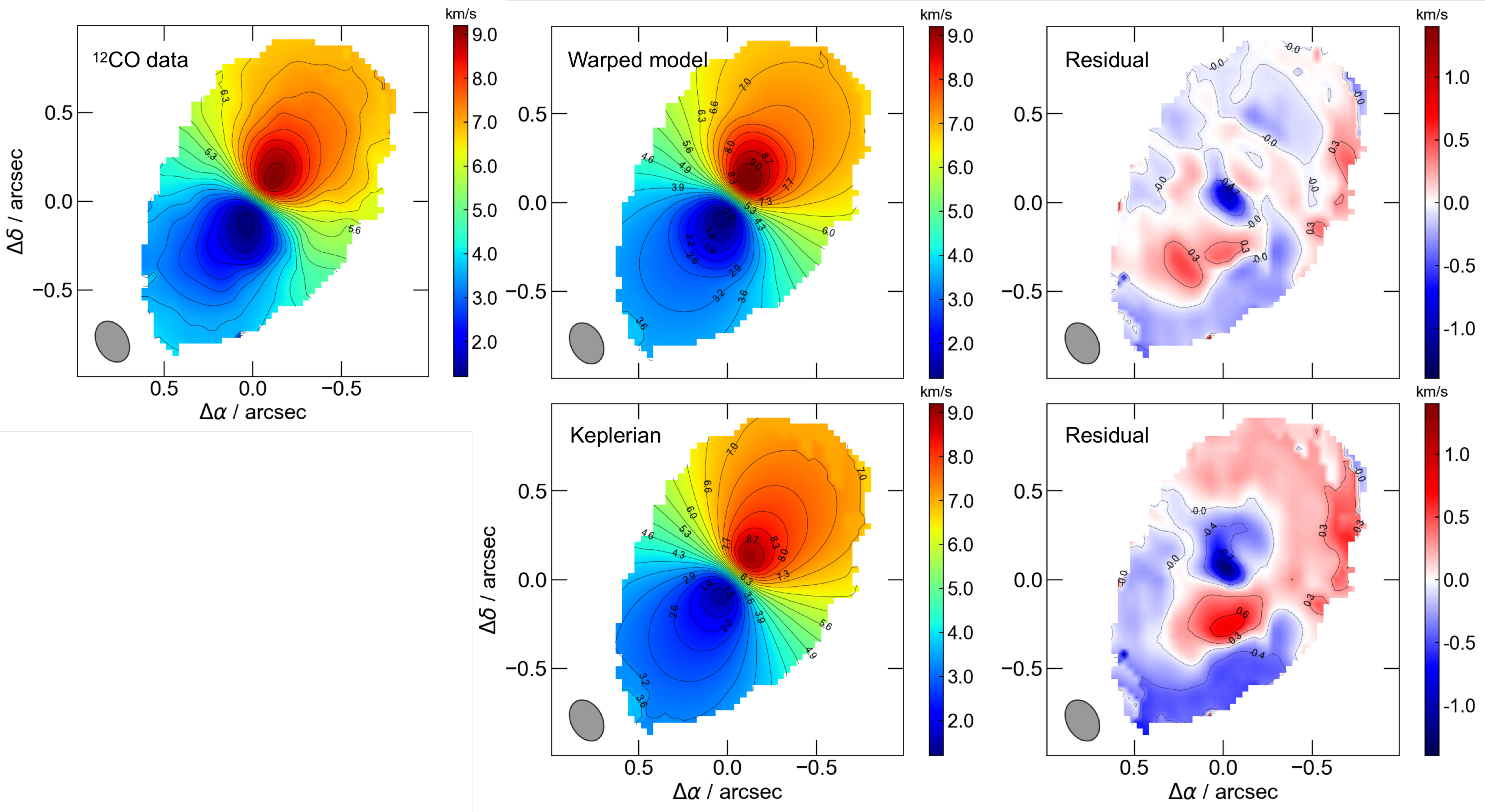}
            \caption{Best fit model for the intensity-weighted velocity field of the \element[ ][12]{CO} emission in the disk (middle panels) for a Keplerian disk (bottom row) and a warped disk (top row). Note that we only compare the velocity field in those regions where the CO emission is above a 5$\sigma$ threshold in the observed moment 0 map. We show the observed intensity-weighted velocity field  in the top left panel, and the residual after subtracting the model from the observations in the right panels.}
            \label{fig:co_model2}
        \end{figure*}

        We compare the observed moment 1 map both with a purely Keplerian disk and with our best-fit solution for a disk warp in Figure \ref{fig:co_model2}. The velocity residuals show that the Keplerian model (bottom right panel) cannot account for the velocity field in the inner region, and produces red and blue residuals that have a different PA from the outer disk.  A mildly warped disk (represented here by an inner region with different PA and inclination) yields, as expected, a better fit to the data than the purely Keplerian model. The best fit inclination for the CO outer disk is 10.0\degr~ lower than the value derived from the dust continuum disk (i.e closer to face-on), while the inner disk inclination is halfway between these values. Similarly the best fit PA for the CO outer disk is 11.5\degr~ lower compared to the value derived from the continuum, while the inner disk PA is 5.0\degr~lower compared to the PA of the dust disk.  The most significant residual after subtracting the best fit warp model is approximately at the stellar position, where our model overpredicts the beam-averaged velocity by 0.7 km s$^{-1}$ in a region the same size as our beam (i.e., likely unresolved).

	\subsection{Orbital fitting of HD~100453~AB}\label{dis:orbit}

	    In the previous sections we show that a co-planar model for the orbit of HD~100453~B may not succeed as well as first thought to match the observations, and in particular the CO data. \citet{2018ApJ...854..130W} present the most complete set of astrometric data for this system yet and we re-assess the orbit and the assumption of co-planarity starting from the same astrometric data.\\

	    We fit the relative orbit of HD~100453~B with respect to the HD~100453~A disk assuming a Keplerian orbit projected on the plane of the sky.  In this formalism, the astrometric position of the companion can be written as:
	    
	    \begin{align}
		    x &= \Delta Dec = r\left(\cos(\omega + \theta)\cos\Omega - \sin(\omega + \theta)\cos i \sin\Omega \right)\label{xMCMC}\\
		    y &= \Delta Ra = r\left(\cos(\omega + \theta)\sin\Omega + \sin(\omega + \theta)\cos i \cos\Omega \right)\label{yMCMC}
	    \end{align} 
	    
	    where $\Omega$ is the longitude of the ascending node (measured counterclockwise from North), $\omega$ is the argument of the periastron, $i$ is the inclination, $\theta$ is the true anomaly, and $r = a(1-e^2)/(1+e\cos\theta)$ is the radius, where $a$ stands for the semi-major axis and $e$ for the eccentricity.  The orbital fit we performed uses the observed astrometry measurements given in \citet[][Table 2]{2018ApJ...854..130W} to derive probability distributions for elements $P$ (period), $e$, $i$, $\Omega$, $\omega$, and time for periastron passage $t_p$.  Elements $a$ and $P$ can be deduced from one another through Kepler's third law.

	    We used two complementary fitting methods, as described in \citet{2012A&A...542A..41C}: (i) a least squares Levenberg-Marquardt (LSLM) algorithm to search for the model with the minimal reduced $\chi^2$, and (ii) a more robust statistical approach using the Markov-Chain Monte Carlo (MCMC) Bayesian analysis technique \citep{2005AJ....129.1706F,2006ApJ...642..505F} to probe the distribution of the orbital elements.  Ten chains of orbital solutions were conducted in parallel, and we used the Gelman-Rubin statistics as a convergence criterion \citep[see][for details]{2006ApJ...642..505F}.  We picked randomly a sample of 500,000 orbits into those chains following the convergence.  This sample is assumed to be representative of the probability (posterior) distribution of the orbital elements, for the given priors.  We chose the priors to be uniform in x = $(\ln P,e,\cos i,\Omega+\omega,\omega-\Omega,t_p)$ following \citet{2006ApJ...642..505F}.  As explained therein, for any orbital solution, the couples ($\omega$,$\Omega$) and ($\omega + \pi$,$\Omega + \pi$) yield the same astrometric data, this is why the algorithm fits $\Omega+\omega$ and $\omega-\Omega$, which are not affected by this degeneracy.  The system distance and total mass used for the fitting are 103 pc and 1.9 M$_\odot$.\\
		
	    We calculate the relative inclination between the orbit of HD~100453~B and the HD~100453~A + disk system using the longitude of node $\Omega$ (equivalent to the PA for disks) which is the angle of the intersection line between the disk and sky plane, measured from the North, and the inclination i, which is the angle between the disk and sky planes. The relative inclination between two planes depends on i$_1$ and i$_2$, but also on the difference $\Omega_1$ - $\Omega_2$ following: 
	    
	 \begin{equation} \label{eq3}
		    \cos i_r = \cos i_1 \cos i_2 + \sin i_1 \sin i_2 \cos (\Omega_1 - \Omega_2).
        \end{equation}
        
\begin{figure*}
            \centering
            \includegraphics[width=\hsize]{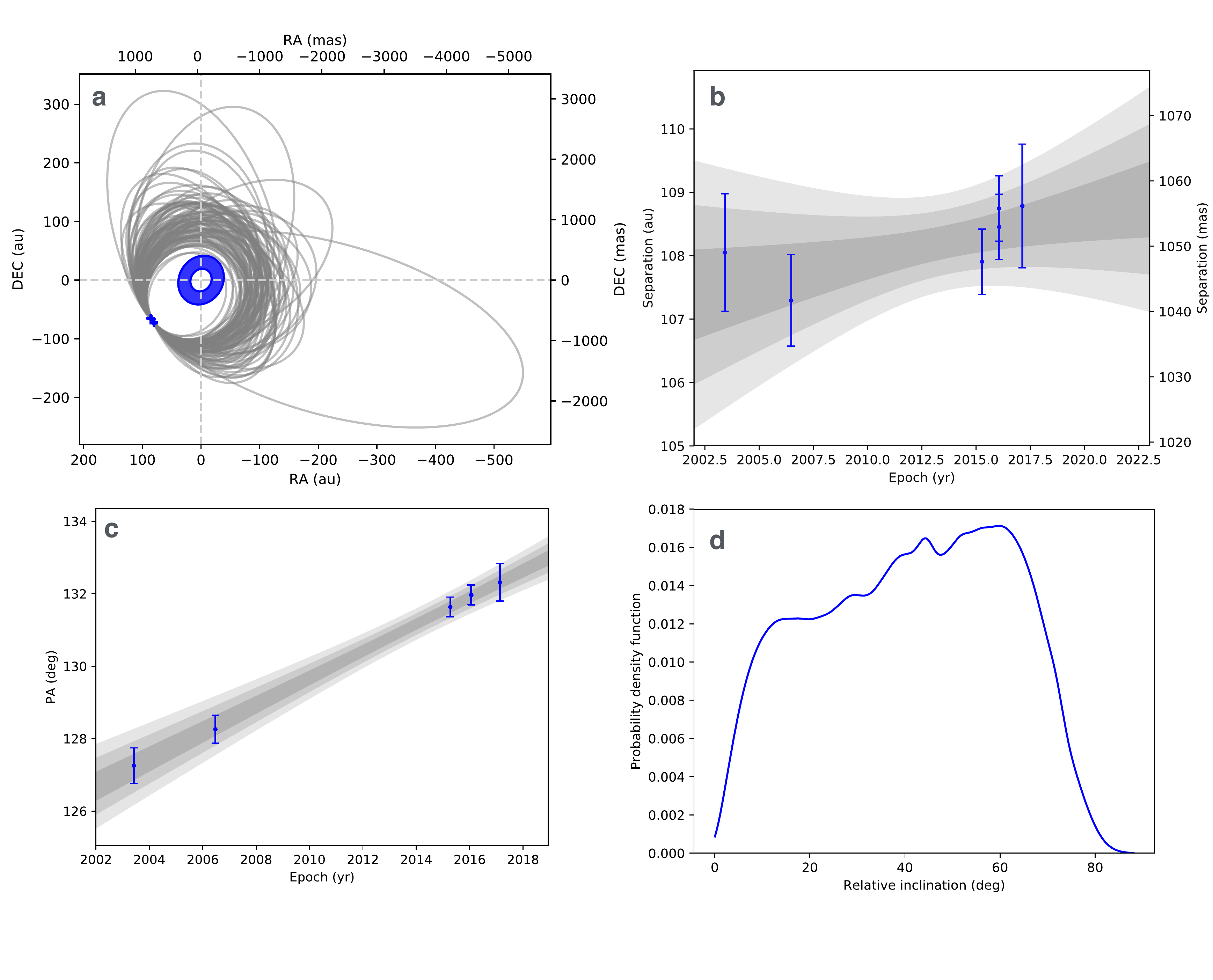}
            \caption{Summary of the orbital fitting results. \textbf{a:} Plots of a hundred trajectories obtained with the MCMC algorithm for the orbit of HD~100453~B. A cartoon of the dust disk is shown at the center. \textbf{b:} Evolution of separation with respect to time. The three shades of grey represent the 1, 2 and 3 $\sigma$ intervals. \textbf{c:} Similar to the 2nd panel, but for the evolution of position angle with respect to time. \textbf{d:} Posterior distribution of the relative inclination between the HD~100453~B orbit and the disk plane.}
            \label{fig:orbit}
        \end{figure*}

	    Despite that the astrometric measurements only cover a small fraction of the orbit, we obtain a consistent fit to the orbit with $\chi _r ^2$ values between 0.5 and 2. A sample of the best-fit orbits is shown in Figure \ref{fig:orbit}, the corner plot showing the posteriors for the orbital fitting in Figure \ref{fig:posteriors}. \gp{The previously mentioned inherent ambiguity of direct imaging regarding the couple ($\Omega$,$\omega$) induces a bimodal posterior distribution for these two parameters. Radial velocity data are needed in order to remove the degeneracy.  On the other hand, the loosely constrained and probably low eccentricity prevents a robust determination of the argument of periastron and the periastron passage. A longer orbital coverage would be necessary to resolve a clear curvature in the orbit and further constrain all the orbital elements.}\\
	
	    We are able to reasonably constrain the semi-major axis of the orbit to be close to the projected value and the eccentricity to be low. Our results are mostly in agreement with the results presented in \citet{2018ApJ...854..130W} but with two deviations. \textbf{Firstly}, whereas \citet{2018ApJ...854..130W} conclude that the inclination of the companion's orbit is co-planar with the disk to within a few $\sigma$, our calculations indicate that a co-planar orbit is not favoured with a most likely relative inclination of 60\degr~(right panel of Figure \ref{fig:orbit}). This is most likely because \citet{2018ApJ...854..130W} did not account for the longitude of node in their determination of the relative inclination. \textbf{Secondly}, the likelihood for the orbital eccentricity of the companion in our calculation peaks at zero eccentricity and it is safe to assume the orbit is bound as the probability distribution of the eccentricity rules out solutions with an eccentricity higher than 0.5 at a 97\% probability. This value is lower than that found by \citet{2018ApJ...854..130W} who find a probability distribution that peaks between values of 0.1 and 0.2.

\section{Discussion}\label{sec:discussion}

	In this Section we tie together our observations with the outcomes of the analysis and discuss the most likely orbit for HD~100453~B, the origin of the detected spiral arms, and the implications thereof on the origin of the disk inner cavity.

	\subsection{The orbit of HD~100453~B}\label{res:synth}

	    Given the current evidence we deem it unlikely that the double armed spiral pattern in this system is excited by an external companion in a close-to co-planar orbit as previously suggested. Even though the outer edge of the dust disk extends to a radius in agreement with tidal truncation by such a companion, the gas disk is not. This disk, as traced by CO, extends out up to a distance greater than the projected separation of the companion. We also do not detect emission from the circumsecondary disk.
	    	    
	    To test for the influence of tidal truncation on the gas in the disk we simulated a system similar to HD~100453.  These simulations show that the material that was originally outside the primary's Roche lobe is captured into a circumsecondary disk or ejected onto a circumbinary ring.	 After a fast initial redistribution of disk material the system continues to evolve on a viscous timescale. This timescale is shorter for the smaller circumsecondary disk, possibly explaining its non-detection.  The circumbinary disk is more significant in our simulations and it is possible that it would survive even up to the current age of the system. The properties of this disk heavily depend on the companion orbit and such a circumbinary disk may even not be present for significantly misaligned orbits. This needs to be tested with future simulations. \\
		
	    A misaligned orbit for HD~100453~B would explain the large extent of the observed circumprimary CO disk as for such orbits the tidal torque on the disk reduces with a factor of $\approx$ cos$^\textrm{8}$(\textit{i}) for misalignment angle \textit{i} \citep{2015ApJ...800...96L}. A secondary on a misaligned orbit can of course also be comfortably outside the primary's Roche lobe while its projected location is close to or overlapping with the disk edge. 
	    	
	    The observed CO disk does show signs of dynamical disturbance through the warped circumprimary disk and through the more elongated spatial distribution of the \element[ ][12]{CO} emission compared to the dust disk geometry and the rarer CO isotopologues. In our SPH simulations small amounts of gas fill the Roche lobe of the primary which closely mimics the \gpp{more stretched out \element[ ][12]{CO} disk} (c.f. the 2 and 30 $\sigma$ contours of the model CO emission shown in the bottom left panel of Figure \ref{fig:sph_sim1}).  This more elongated structure for the \element[ ][12]{CO} emission is in qualitative agreement with the distribution of CO gas in our simulations and we interpret it as a reservoir of lower-density material which is distributed along the major axis by tidal interactions between the gas disk and the companion. Connecting the warped CO disk to HD~100453~B also hints at an inclination for the companion orbit that is closer to face-on than that of the circumprimary disk because the inclinations derived for the inner and outer disk are progressively closer to face-on compared to the inclination calculated from the (midplane) dust emission (c.f. Section \ref{sec:rt}). %

	    Our orbital fitting shows that while we cannot constrain the relative inclination between the companion and the disk, a co-planar orbit is not favoured. Rather, the probability density function for the relative inclination peaks at a misalignment of $\approx$60\degr. Two quantities that we can reasonably constrain are a low eccentricity orbit and a semi-major axis close to the projected separation.\\
	    	
	    Finally, the south-western spiral pointing towards the companion is expected to contain more mass if an external companion on a co-planar orbit was to excite the double armed spiral.  We find instead a mm counterpart to the northern spiral suggesting that this is the primary spiral arm.\\

	    Given the above, we re-evaluate the causal connection between the external companion and the double spiral arms. The CO disk does show signs of tidal disturbance and while the orbit of the companion is of low eccentricity, it most likely is significantly misaligned compared to the plane of the disk. Such a misaligned orbit allows for weaker truncation of the circumprimary disk and explains both the warped outer disk and the CO emission that is seen up to distances similar to the separation of the companion.  
	
	    \gpp{Such} a companion that orbits in a plane that is misaligned compared to the disk could still excite double spiral arms, but it is as of yet unclear what those spiral arms would look like in terms of opening angle and surface density contrast. We therefore consider alternative scenarios that could also generate the observed spiral pattern in section \ref{sec:discussion_origin}.

    \subsection{The disk cavity + misaligned inner disk}

        We resolve an inner cavity extending up to 23 au \gp{from the mm dust continuum emission} that contains an unresolved mm counterpart to the small misaligned inner disk previously detected using near- and mid-IR interferometric observations \citep{2015A&A...581A.107M, 2017A&A...599A..85L}, and whose presence is corroborated by two shadows cast on the outer disk \citep{2017A&A...597A..42B}.  The spectral index we determine for the inner disk from the limited 0.1 mm bandwidth available is 3.5 $\pm$ 1.3.  This is consistent with emission originating from a dusty disk.  The NIR emission from the inner disk is best fit with a Gaussian with an inclination of 48\degr~and a PA of 80\degr~\citep{2017A&A...599A..85L}, significantly misaligned with respect to the values we determine for the circumprimary disk. 
        The size of the cavity detected in scattered light is $\approx$ 21 au \citep{2015ApJ...813L...2W}, comparable to the size of the cavity in mm emission.  The geometry of the outer cavity wall deviates at both wavelengths from circular symmetry and has a more hexagonal shape.         
        
        Secular precession resonances in young binary systems with mass ratios of the order of 0.1 can generate large misalignments between the circumstellar disk and a companion \citep{2017MNRAS.469.2834O}, and these authors suggest that the misalignment seen in HD~100453 could have been generated by resonance crossing and that such a scenario implies that a low-mass (between $\approx$0.01 and 0.1 M$_\odot$) companion is residing inside the cavity with an orbit that is aligned with the outer disk. Such a companion in a circular orbit would need to orbit at $\approx$ 13 au to truncate the circumbinary\footnote{We refer to this disk as the circumprimary disk in the rest of this manuscript because, while likely, the presence of a companion inside the cavity has not been confirmed by direct obesrvations.} disk at 23 au \citep{1994ApJ...421..651A}. \gp{More eccentric orbits would allow for values smaller than 13 au for the companion orbit}.

        It is interesting to note that the HD~100453 system shares many similarities with the much better studied HD~142527 system, such as a small and misaligned inner disk, a large disk cavity, spiral arms and shadows cast by a misaligned inner disk detected in scattered light, and azimuthally asymmetric mm emission in the outer disk.  Recent work by \citet{2018MNRAS.477.1270P} shows that the interaction of a companion inside the cavity on an inclined and eccentric orbit can reproduce the spirals, shadows, and horseshoe geometry of dust emission detected in that disk, as well as a non-circular geometry of the outer cavity wall.
        
        The presence of such a close-in companion in this system is supported by the lack of detected CO emission from the HD~100453 disk within 7 au.  Furthermore, the lower-than expected observed velocities of the CO gas close to the stellar position ($\approx$ 10\% to 20\% lower than the local Keplerian velocity, c.f. Figure \ref{fig:co_model2}) is consistent with a velocity signature left by a close-in companion. \citet{2018MNRAS.480L..12P} show that the spiral wakes left by these bodies imprint asymmetric velocity patterns, where the maximum deviation from Keplerian rotation occurs at the outer spiral wake launched by a giant planet. Once these kinematic signatures get convolved with our beam they would appear similar to the deviation we detect. \\
        
	    These lines of reasoning all point towards the presence of a companion inside the cavity. Determining the precise properties of such a companion requires more data and investigations and is outside the scope of this work, but the constraints on the orbital parameters and mass are that it should be able to drive the misalignment of the inner disk while not leaving a gravitational fingerprint on the velocity field of the CO gas in the cavity that would have stood out in our observations.
                        
	\subsection{Possible origins of the spiral arms}\label{sec:discussion_origin}
	
        \citet{2015MNRAS.451.1147J} argue that planet-induced spiral arms are unlikely to be detected with current instruments, and suggest that all as of yet observed spiral arms are instead pressure scaleheight perturbations. Together with the relative brightness of the northern spiral arm and the extent of the CO disk this motivates us to explore alternative origins for the spiral arms. 
	    
        Self-gravity can cause parts of the disk to collapse and form spiral arms in the process if the disk is sufficiently massive. Typically a disk needs to hold $\approx$ 10\% of the mass of the central star for gravitational instabilities to become relevant \citep[see e.g. the review by][]{2016ARA&A..54..271K}, a condition that is far from fulfilled in the disk around HD~100453~A.  A gravitationally unstable disk is unlikely to be the cause for the detected spiral arms. 
        
        Stellar fly-by scenarios can, under certain circumstances, also generate two near-symmetric spiral arms in disks \citep{2003ApJ...592..986P}, but the low eccentricity of the orbit of HD~100453~B indicates the companion is on a bound orbit which makes a recent fly-by an equally unlikely candidate for provoking the spiral arms. 
        	
        A companion inside the disk cavity could drive a slow precession of the misaligned inner disk. If the direction of this precession is prograde and a region of the outer disk rotates at the same frequency as the shadow cast by this precessing inner disk  ($\approx$ 85 years for a launching location for the spirals of 0\farcs22), spiral arms whose pitch angle much resemble those caused by a planet can develop at the location of the shadow \citep{2016ApJ...823L...8M, 2018MNRAS.tmpL...3M}. Slight asymmetries in the tilted inner disk affect the depth of shadows and thus the relative strength of the spiral arms. A weaker shadow on the western disk then would be able to explain the non-detection of a mm counterpart to the southern spiral arm.\\

\section{Conclusions}\label{sec:conclusion}

    We resolve the disk around HD~100453~A into a disk of dust continuum emission between 23 and 41 au,  an unresolved inner disk, \gp{and excess mm emission at the location of the northern spiral arm  detected using scattered light imaging. Two likely origins for this excess emission are [1] that it is a mm counterpart to the spiral arm, or [2], that it is a narrow vortex associated with the spiral arm either through having a common origin or by inducing the spiral arm.  We do not detect emission from the location of HD~100453~B and put a 3$\sigma$ upper limit on the dust content for that disk of 0.03 Earth masses}.  The CO emission from the circumprimary disk extends out to 1\farcs10 and shows a velocity pattern that is mostly Keplerian but with a 10\degr~warp between inner and outer disk.  The morphology of the $^{12}$CO disk is more elongated along the major axis when compared to the $^{13}$CO, C$^{18}$O, and mm dust emission, likely as a consequence of tidal disruption of the circumprimary disk by HD~100453~B.\\

	Our fit to the orbit of HD~100453~B suggests a significantly misaligned orbit w.r.t. the circumprimary disk. Such an orbit is supported by our SPH simulations, which show that a companion on a co-planar orbit cannot reproduce the detected spatial extent of the CO disk nor our detection of mm emission from the northern spiral arm.  It is possible that a companion at larger separation and/or on an inclined orbit reproduces the morphology of the detected CO emission better but it is as of yet unclear if a companion on a sufficiently misaligned orbit can qualitatively reproduce the spiral arm morphology of this system. Pending detailed calculations of the impacts by a significantly inclined orbit of the companion on the circumprimary disk we suggest an alternative scenario that could also generate the observed spiral pattern.

	Given the relatively low mass of the disk and the low eccentricity of the orbit of HD~100453~B we deem a recent fly-by or a gravitational instability in the disk unlikely to provoke the spiral arms. Instead, we suggest that co-moving shadows of a precessing inner disk as possible cause for the detected spiral arms.  A small misaligned inner disk has been detected using near infrared interferometry and its shadows are visible on the outer disk at roughly the same location as the launching points of the spiral arms.  Such a misaligned inner disk, the non-detection of CO emission from the inner 7 au, and the 23 au large cavity in the dust disk, all can be explained by a companion inside the disk cavity orbiting at a distance between a few and $\approx$ 13 au. 
	
	\gp{All features described in this manuscript are illustrated in Figure \ref{fig:overview} together with a list of relevant figures in which they are visible.}

\begin{acknowledgements}
    This paper makes use of data from ALMA programme 2015.1.00192.S.  ALMA is a partnership of ESO (representing its member states), NSF (USA) and NINS (Japan), together with NRC (Canada) and NSC and ASIAA (Taiwan), in cooperation with the Republic of Chile.  The Joint ALMA Observatory is operated by ESO, AUI/NRAO and NAOJ.  The National Radio Astronomy Observatory is a facility of the National Science Foundation operated under cooperative agreement by Associated Universities, Inc.  GP, FM and JFG acknowledge funding from ANR of France under contract number ANR-16-CE31-0013 (Planet-Forming-Disks), LR acknowledges funding from ANR of France under contract number ANR-14-CE33-0018 (GIPSE). JFG thanks the LABEX Lyon Institute of Origins (ANR-10-LABX-0066) of the Universit\'e de Lyon for its financial support within the programme ‘Investissements d’Avenir’ (ANR-11-IDEX-0007) of the French government operated by the ANR. LC acknowledges support from FONDECYT grant 1171246, and SC acknowledges support from FONDECYT grant 1130949 and from the Millennium Science Initiative (Chilean Ministry of Economy) through grant RC130007.  This project is supported by CNRS, the OSUG$@$2020 labex and the Programme National de Planetologie (PNP, INSU) and Programme National de Physique Stellaire (PNPS, INSU). The orbital fit presented in this paper was performed using the Froggy platform of the CIMENT infrastructure (https://ciment.ujf-grenoble.fr), which is supported by the Rhone-Alpes region (GRANT CPER07 13 CIRA), the OSUG$@$2020 labex (reference ANR10 LABX56) and the Equipe$@$Meso project (reference ANR-10-EQPX-29-01) of the programme Investissements d'Avenir, supervised by the Agence Nationale pour la Recherche. SPH simulations were run at the Common Computing Facility (CCF) of LABEX LIO.
\end{acknowledgements}


\begin{appendix}\label{appendix:A}

    \section{Orbital fitting}

	\begin{figure*}
            	\centering
            	\includegraphics[width=\hsize]{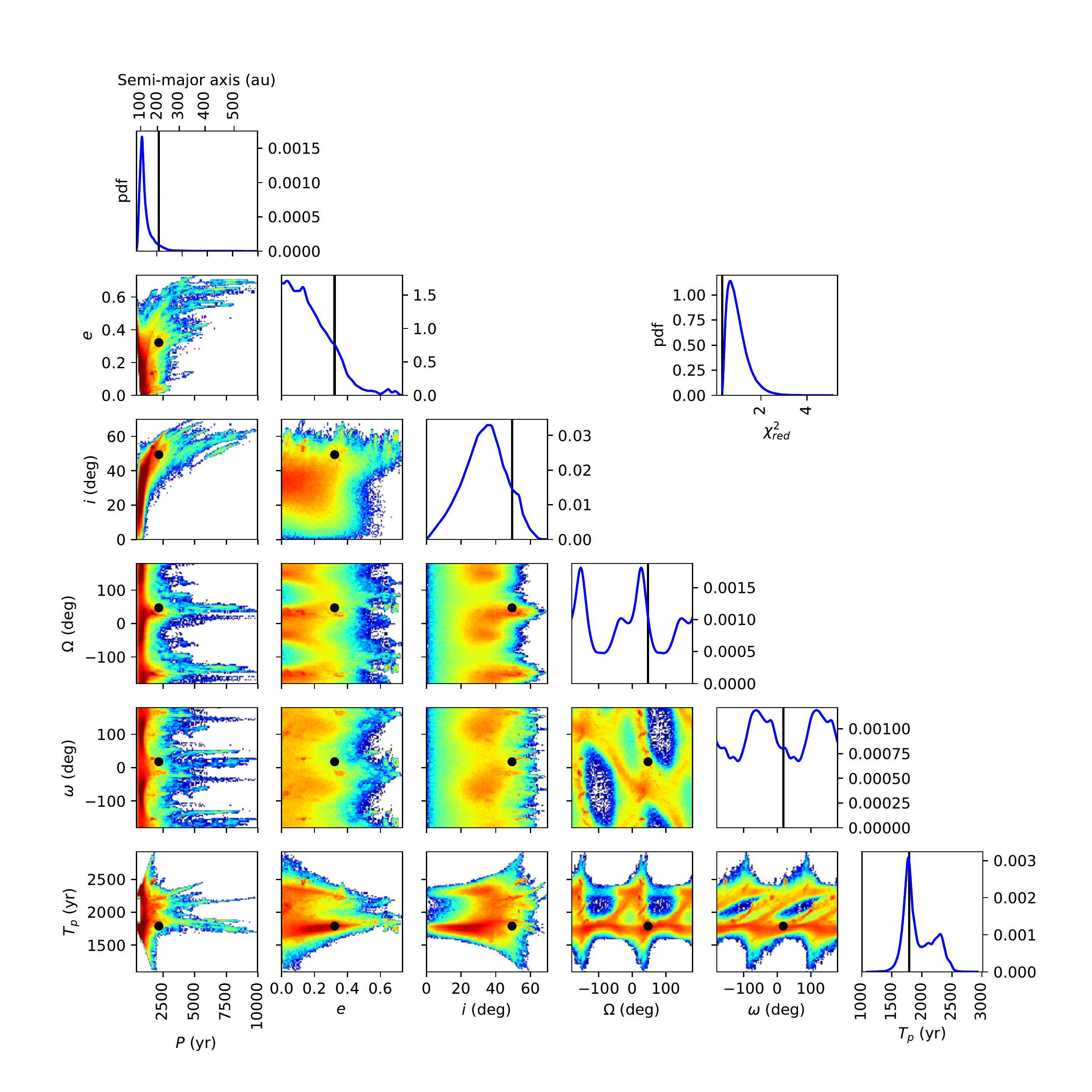}
            	\caption{Distribution and correlations of each of the orbital element fitted by the MCMC algorithm. The black lines and points depict the best fitting orbit (better $\chi^2$), obtained with the LSLM algorithm. The color scale is logarithmic, blue corresponds to 1 orbit and red to 1000}
            	\label{fig:posteriors}
    \end{figure*}

    \section{Summary Figure}

	\begin{figure*}
            	\centering
            	\includegraphics[width=\hsize]{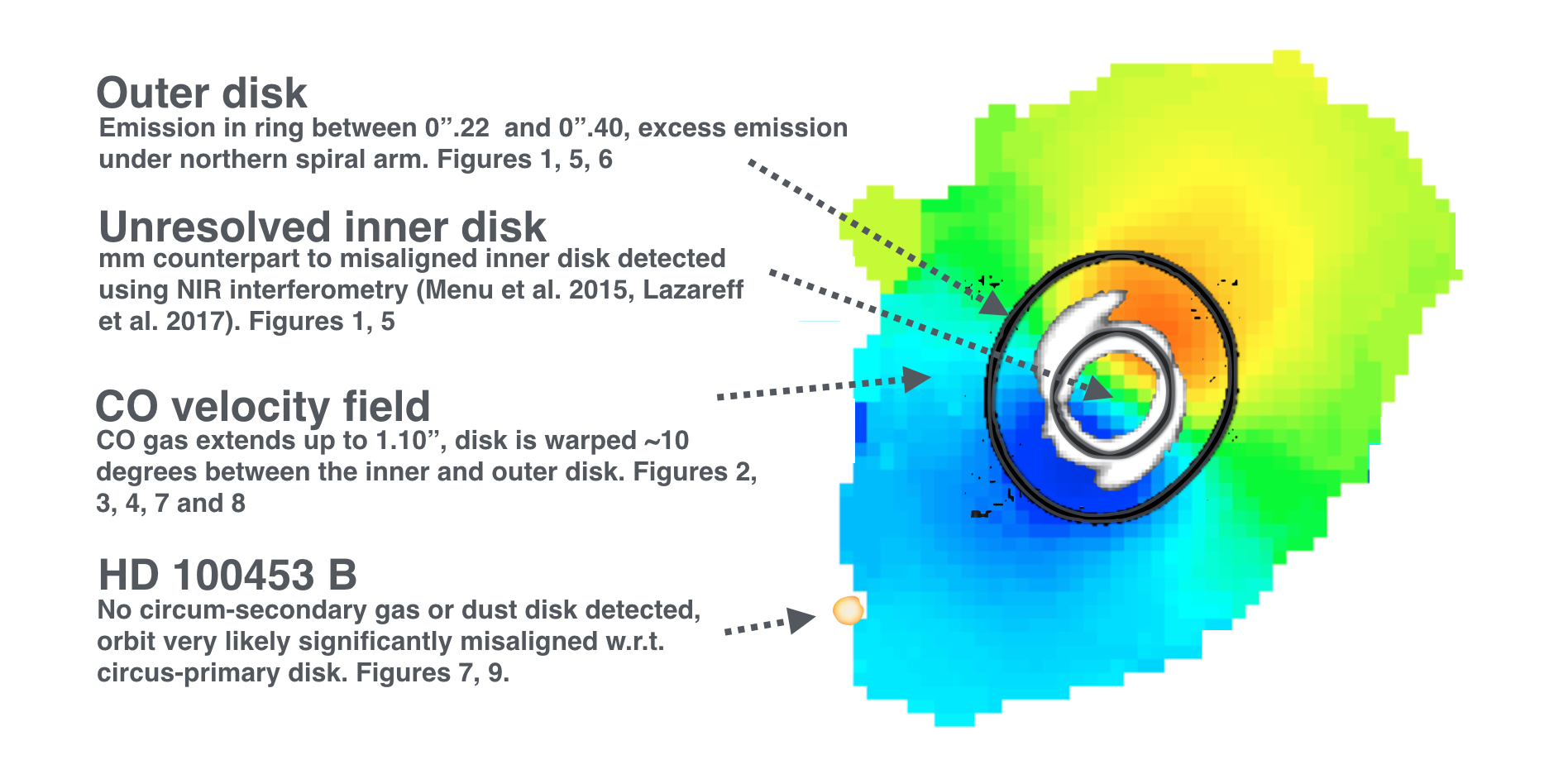}
            	\caption{Summary of all features of the HD~100453 system discussed in this paper made using Figures \ref{fig:continuum} and \ref{fig:moments-12co21-100453} presented in this manuscript, together with a list of Figures relevant to that feature. We also show an overlay of the  J-band Q$_\phi$ image reproduced from Figure 2 in \citet{2017A&A...597A..42B}.}
            	\label{fig:overview}
    \end{figure*}

\end{appendix}

\end{document}